\documentclass[a4paper,12pt]{revtex4}
\usepackage{amssymb}
\usepackage{amsmath}
\usepackage{graphics}
\usepackage{color}
\usepackage[T1]{fontenc}
\usepackage{times,mathptm}
\usepackage[cp1250]{inputenc}
\usepackage{epsfig,epsf}
\newcommand{\be}{\begin{equation}}
\newcommand{\ee}{\end{equation}}

\newcommand\beq{\begin{eqnarray}}
\newcommand\eeq{\end{eqnarray}}

\begin{document}

\title{Evidence for breakdown of the DGLAP description in diffractive DIS at HERA}

\author{M. Sadzikowski\,\footnote{Talk presented during the conference DIS2012, 
Bonn, 2012.}, L. Motyka, W. Slominski}

\affiliation{Smoluchowski Institute of Physics, Jagiellonian
University, Reymonta 4, 30-059 Krak\'ow, Poland}



\begin{abstract}
 HERA data on diffractive DIS show deviations from twist 2 DGLAP predictions below
$Q^2\sim 5$ GeV$^2$ at low pomeron $\xi$, which may reach up to 100 $\%$. These deviations are consistent
with higher twists effects extracted from the saturation model. It is a first direct
evidence for the higher twists in DIS. This finding affects determination of the
diffractive parton densities that are used for the predictions at the LHC.
\end{abstract}

\maketitle

\section{Introduction}

A large fraction of the deep inelastic scattering (DIS) HERA data are produced due to the processes of diffractive dissociation of
virtual photons $\gamma^\ast p\rightarrow Xp$ (DDIS). In those reactions, viewed from the proton rest frame, the virtual photon $\gamma^\ast$ fluctuates
into the strongly interacting debris which scatter off the proton target and eventually create a final hadronic state $X$. The proton
target remains intact. The conventional description of DDIS is based on the leading twist DGLAP evolution equations which characterize the
QCD hard scale dependence of the diffractive parton distribution functions (DPDFs). This approach is justified by Collins factorization theorem
\cite{collins}. Despite of clear success such description faces an important limitation that follows from neglecting
higher twists. The higher twists contribution becomes relevant below some energy scale which depends on the process. In the
case of the inclusive DIS, a leading twist description of the data is reasonable down to the photon virtuality $Q^2\sim 1$ GeV$^2$ \cite{bartels},
however, in the case of DDIS such description breaks down already at higher scale $Q^2\sim 5$ GeV$^2$. In this presentation
we would like to show, that the deviations of the extrapolated DGLAP description from the DDIS date at low $Q^2$ are consistent with
emergence of the higher twists contribution. This provides a first evidence of the higher twists effects in DIS data and opens
a new window for studying the physics beyond leading twist, both at the experimental and theoretical level.

\section{Cross section and the breakdown of twist-2 description}

\begin{figure}
\centerline{
\includegraphics[width=0.35\columnwidth]{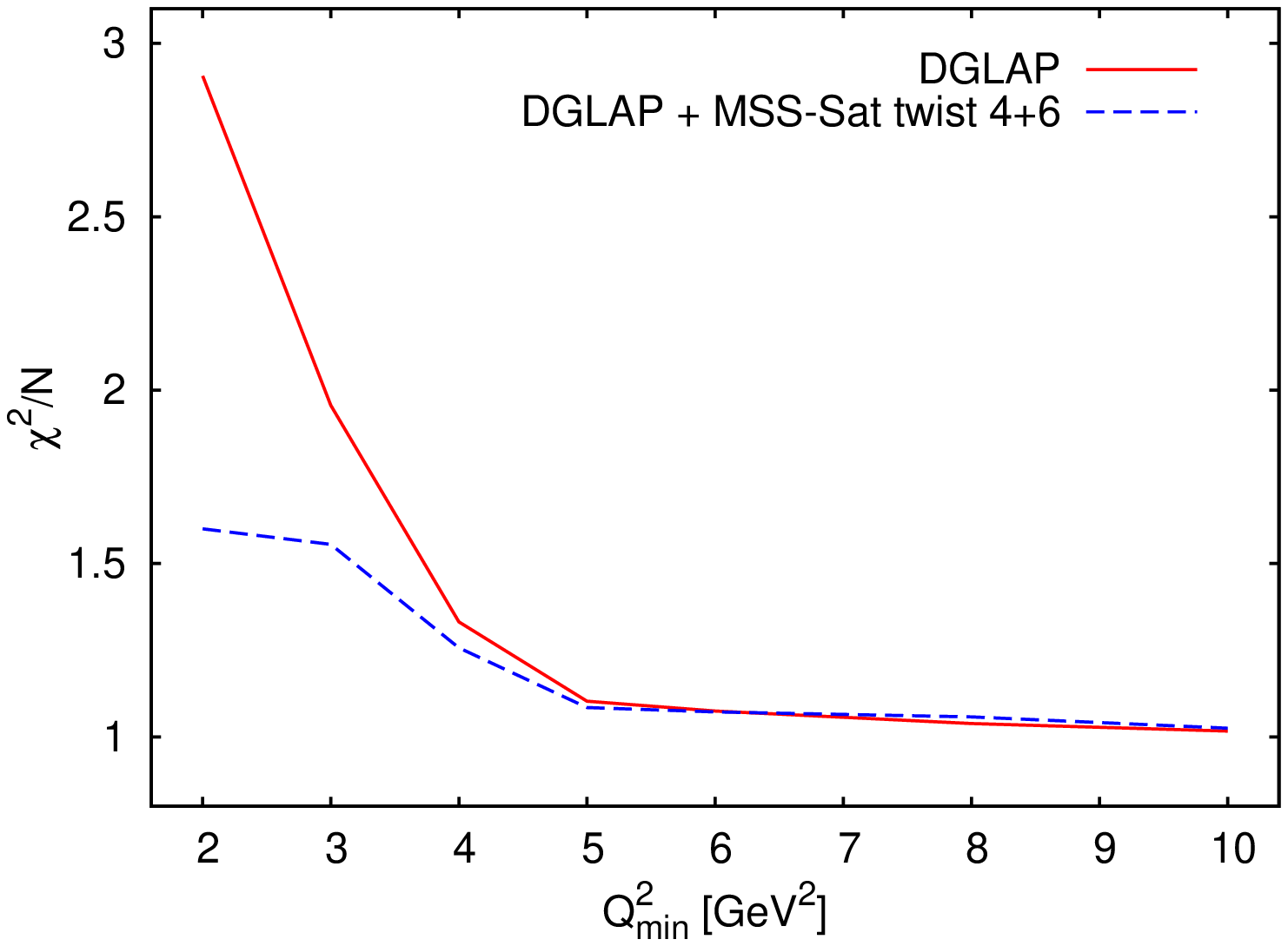}\hspace*{1cm}\includegraphics[width=0.4\columnwidth]{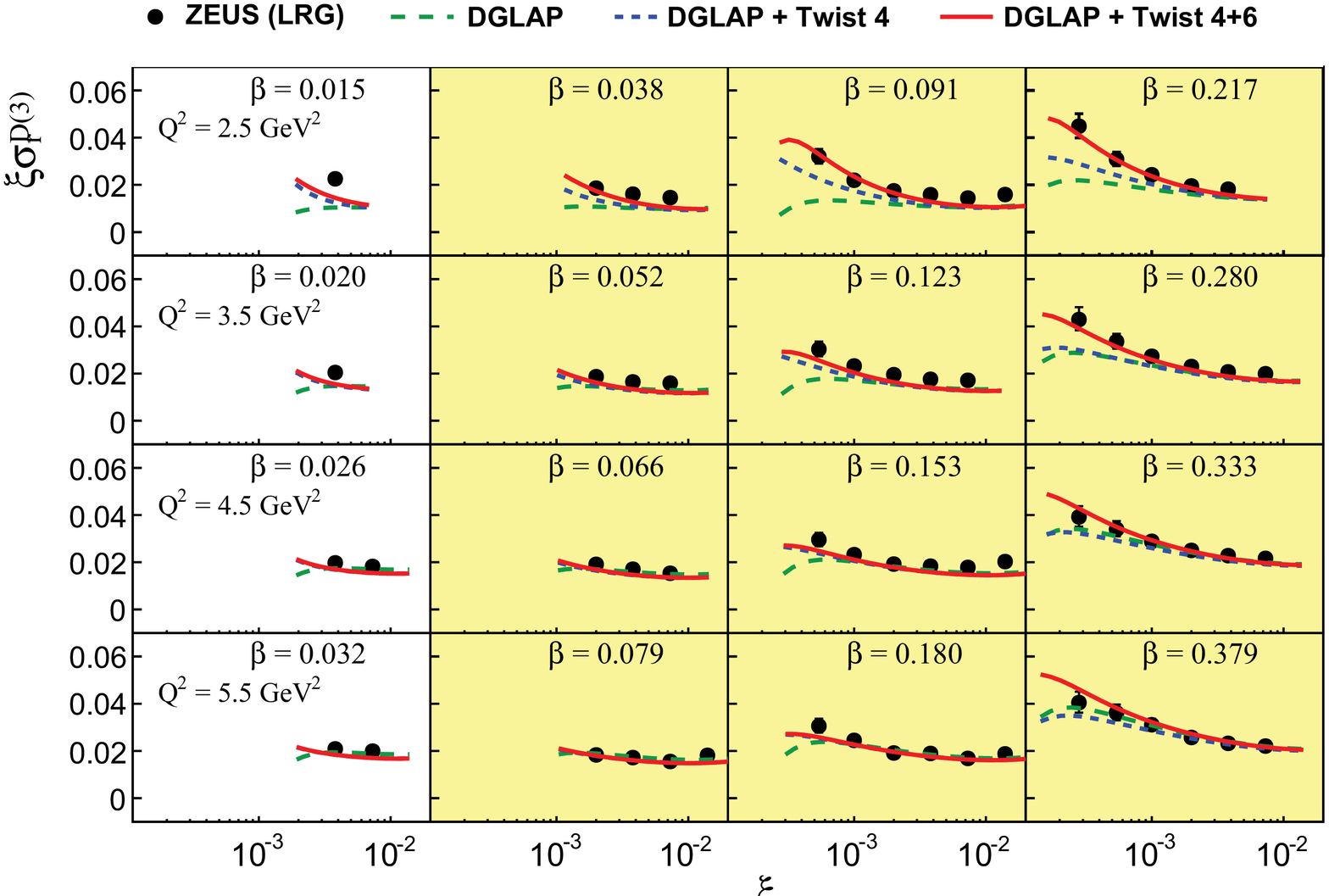}\vspace{-1em}
}
\caption{Left panel - the $\chi^2/\,{\rm d.o.f.}$ for NLO DGLAP and NLO DGLAP + HT fits to ZEUS LRG data \cite{ZEUS} with $Q^2 > Q^2_{\mathrm{min}}$.
Right panel - the LRG ZEUS data for $\xi\sigma_r ^{D(3)}$ at low $Q^2$ compared to a DGLAP fit \cite{ZEUS} and the DGLAP fit with included twist-4 and twist-4 and~6 corrections from the saturation model. In yellow (gray) --- the region of $\beta$ where the correction due to $q\bar q gg$ may be neglected.
\label{fig1}
}
\end{figure}

The DDIS is an quasi-elastic electron - proton scattering process $e(k)p(P)\rightarrow e(k^\prime)p(P^\prime)X(P_X)$ in which the
final hadronic state $X$ with four-momentum $P_X$ is separated in rapidity from the proton, that scatters elastically.
The $t$-integrated ep cross-section reads:
\be
\frac{d\sigma}{d\beta dQ^2 d\xi} = \frac{2\pi\alpha^2_{\mathrm{em}}}{\beta Q^4}[1+(1-y)^2]\sigma_r^{D(3)}(\beta, Q^2,\xi)
\ee
where the invariants read $y=(kq)/(kP)$, $Q^2=-q^2$, $\xi = (Q^2+M_X^2)/(W^2+Q^2)$ and $t=(P^\prime - P)^2$. The
quantity $W^2=(P+q)^2$ is the invariant mass squared in photon-proton scattering, and $M_X^2$ is the invariant mass
of the hadronic state $X$. The reduced-cross-section may be expressed in terms of the diffractive structure functions
$\sigma_r^{D(3)}(\beta, Q^2,\xi)=F_L^{D(3)}+F_T^{D(3)}$, whereas the structure functions $T,L$ may be, respectively, expressed through transversally and longitudinally polarized $\gamma^\ast$ - proton cross sections
$F_{L,T}^{D(3)} = (Q^4/4\pi^2\alpha_{em}\beta\xi)\, d\sigma^{\gamma^\ast p}_{L,T}/dM^2_X$.

In recent analysis \cite{ZEUS} the ZEUS diffractive data were fitted within NLO DGLAP approximation. A satisfactory
good description was found only for $Q^2>Q^2_{min}=5$ GeV$^2$. The ZEUS fits were performed
above $Q^2_{min}$ and then extrapolated to lower photon virtualities. The deviations of the fits rapidly grow with
decreasing $\xi$ and $Q^2$ reaching 100 percent effect at the minimal $Q^2=2.5$ GeV$^2$ and $\xi\simeq 4\cdot 10^{-4}$.
We confirmed this result calculated $\chi^2/$d.o.f. for subsets of ZEUS LRG data with $Q^2>Q^2_{min}$ and
$\beta > 0.035$ \cite{MSS} (see Fig. 1, left panel). The cut off in $\beta$ serves to reject contributions from highly resolved projectiles.
It is clear from this discussion that the leading twist DGLAP evolution is unable to describe the DDIS data below
$Q^2\simeq 5$ GeV$^2$ and at the low~$\xi$.

\section{Estimation of the higher twist contributions}

The large energy limit of the DDIS scattering may be described within the framework of the
colour dipole model \cite{NZ, GBW}. In this approach the $\gamma^\ast p$ process is factorized
into an amplitude of photon fluctuation into the partonic debris and then scattering of these
states off the proton by the multiple gluon exchange. We take into account the contributions
from the fluctuation of the photon into a colour singlet quark-antiquark pair $q\bar{q}$ and
into $q\bar{q}$-gluon triple. We neglect the contribution from the higher Fock states.  This
gives the $t$-integrated $\gamma^\ast p$ cross section
$d\sigma_{L,T}^{\gamma^\ast p}/dM_X^2 = d\sigma_{L,T}^{q\bar{q}}/dM_X^2+d\sigma_{L,T}^{q\bar{q}g}/dM_X^2$.
Assuming an exponential $t$-dependence of diffractive cross-section, one finds for the $q\bar{q}$ component
\be
\label{diff_cross_sec_qq}
\frac{d\sigma_{L,T}^{q\bar{q}}}{dM_x^2} = \frac{1}{16\pi b_D}\int\frac{d^2p}{(2\pi)^2}\int_0^1 dz\,\, \delta\left(\frac{p^2}{z\bar{z}}-M_x^2\right)
\sum_f\sum_{spin}\left| \int d^2r e^{i\vec{p}\cdot\vec{r}}\psi^f_{h\bar{h},\lambda}(Q,z,\vec{r})\sigma_d (r,\xi)\right|^2 .
\ee
where $b_D$ is a diffractive slope, $z\bar{z}=z(1-z)$ and the first sum runs over the three light flavours. The second sum of (\ref{diff_cross_sec_qq}) means summation over massless (anti)quark helicities $(\bar{h}) h$ in the case of longitudinal photons whereas for transverse photons there is an additional average over initial photon polarizations $\lambda$. The squared photon wave functions reads, e.g. \cite{MKW}
\be
\label{photon_wave_function_2}
\sum_{spin}\psi^f_{h\bar{h},\lambda}(Q,z,\vec{r})\psi^{f\,\ast}_{h\bar{h},\lambda}(Q,z,\vec{r}^{\,\prime}) = \frac{N_c\alpha e_f^2}{2\pi^2}
\left\{
\begin{array}{lr}
4 Q^2 (z\bar{z})^2 K_0(\epsilon r) K_0 (\epsilon r^\prime) & \;\; (L) \\
\epsilon^2(z^2+\bar{z}^2) \frac{\vec{r}\cdot\vec{r}^{\,\prime}}{rr^\prime}K_1(\epsilon r)K_1(\epsilon r^\prime) & \;\; (T)
\end{array}\right.
\ee
where $K_0, K_1$ are McDonald-Bessel functions and $\epsilon = \sqrt{z\bar{z}}Q$. We use the GBW parametrization
\cite{GBW} for the dipole-proton cross section $\sigma_d(r,\xi) = \sigma_0(1-\exp(-r^2/4R_\xi^2))$. The saturation radius in DDIS depends
on $\xi$, $R_\xi=(\xi /x_0)^{\lambda /2}$ GeV$^{-1}$ and $\sigma_0=23.03$ mb, $\lambda = 0.288$, $x_0=3.04\cdot 10^{-4}$.

The contribution of the $q\bar{q}g$ component of $\gamma^\ast$ is calculated at $\beta =0$ and in the soft gluon approximation (the
longitudinal momentum carried by a gluon is much lower then carried by the $q\bar{q}$ pair). This approximation
is valid in the crucial region of $M_X^2\gg Q^2$ or $\beta \ll 1$, where the deviations from DGLAP are observed.
The correct $\beta$-dependence is then restored using a method described in \cite{marquet}, with kinematically
accurate calculations of Ref. \cite{wusthoff}. With these approximations one obtains:
\beq
\label{cross_sec_qqg}
\frac{d\sigma^{q\bar{q}g}_{L,T}}{dM_x^2} &=& \frac{1}{16\pi b_D}\frac{N_c\alpha_s}{2\pi^2}\frac{\sigma_0^2}{M_x^2}\int d^2r_{01}
N^2_{q\bar{q}g} (r_{01},\xi)\sum_f\sum_{spin}\int_0^1 dz|\psi^f_{h\bar{h},\lambda}(Q,z,r_{01})|^2, \\\nonumber
N^2_{q\bar{q}g} (r_{01}) &=& \int d^2r_{02}\frac{r_{01}^2}{r_{02}^2 r_{12}^2}\left( N_{02}+N_{12}-N_{02}N_{12}-N_{01} \right)^2
\eeq
where $N_{ij} = N(\vec{r}_j-\vec{r}_i)$, $\vec{r}_{01}, \vec{r}_{02},\vec{r}_{12}=\vec{r}_{02}-\vec{r}_{01}$ denote the relative positions of quark and antiquark $(01)$, quark and gluon $(02)$ in the transverse plain. The form of $N^2_{qqg}$ follows from the Good-Walker picture of the diffractive dissociation of the photon \cite{munier_shoshi}. The factor $1/M_X^2$ is a remnant of the phase space integration under the soft gluon assumption. The twist decomposition of (\ref{cross_sec_qqg}) is performed using the Mellin transform in the $r_{01}$ variable:
\beq
\label{cross_sec_qqg_mellin}
\frac{d\sigma^{q\bar{q}g}_{L,T}}{dM_X^2} &=&
\frac{N_c\sigma_0^2\alpha_s}{32\pi^3 b_D M_X^2}
\int\frac{ds}{2\pi i}\left(\frac{4 Q_0^2}{Q^2}\right)^{-s}\tilde{H}_{L,T} (-s)
\, \tilde N_{qqg}^2(s),
\eeq
The Mellin transform of $N^2_{q\bar{q}g} (r_{01})$ can be done in two steps. First one defines new integrals
\beq
\label{int_I}
I_1&=&\frac{(Q_0^2)^s}{\pi}\int d^2r_{01} (r_{01}^2)^{s-1} \int d^2r_{02}\frac{r_{01}^2}{r_{02}^2 r_{12}^2}\left[\left(N_{02}+N_{12}-N_{02}
N_{12}\right)^2-N_{01}^2 \right] ,\\\nonumber
I_2&=&\frac{(Q_0^2)^s}{\pi}\int d^2r_{01} (r_{01}^2)^{s-1} \int d^2r_{02}\frac{r_{01}^2}{r_{02}^2 r_{12}^2}2N_{01}\left[ N_{02}+N_{12}-N_{02}N_{12}-N_{01} \right] .
\eeq
where $\tilde{N}_{q\bar{q}g}^2(s) = I_1-I_2$. The integral $I_1$ can be performed exactly,
\beq
\label{I_1}
I_1=\pi \, (Q_0 R_\xi)^{2s} \, 2^{1+s}\,(2^{1+s}-1)\, \Gamma (s)
\times \left[H_s - {}_3F_2(1,1,1-s;2,2;-1)s\right], \;\, H_s = \sum_{k=1}^s\frac{1}{k},
\eeq
and for $I_2$ we use the large daughter dipole approximation $r_{02} \gg r_{01}, \vec{r}_{12}\approx \vec{r}_{02}$ and obtain,
\beq
I_2=\, \pi \, (Q_0 R_\xi)^{2s} \, 2^{1+2s} \, \Gamma (s) \,
\left\{ 1-2^{1-s}+3^{-s} + \, \frac{2^{-s}s}{1+s}\left[1 - {}_2 F_1\left( 1+s,1+s;2+s;-\frac{1}{2}\right)\right]\right\} .
\eeq
The twist decomposition follows from (\ref{cross_sec_qqg_mellin}) as a sum over residues at the negative integer values of $s$. Accuracy of this
approximation is at the level of 5 per cent. 

\section{Discussion}

In Fig. 1, right panel we compare selected results with data: the extrapolated DGLAP results, DGLAP plus twist-4, DGLAP plus twist-4 and twist-6.
 The saturation model results are obtained using the original GBW parameters $\lambda$ and $\sigma_0$, and three massless quark
flavours. In our approach we modified the GBW parameter $x_0$ to $\xi_0=2x_0$ in order to account for the difference between
Bjorken $x$ and $\xi$, the variables used in GBW dipole cross-section in DIS and DDIS respectively. We chose $\alpha_s=0.4$
that provides a good description of data. The conclusions from the analysis and from the Fig. 1
are the following: (i) at twist-2 the DGLAP fit and the twist-2 components of the model are reasonably consistent, 
but all fail to describe the data below $Q^2=5$ GeV$^2$ and at low $\xi$; (ii) a combination of the DGLAP fit
and twist-4 and twist-6 components of the model gives a good description of the data at low $Q^2$. Inclusion of
the higher twists terms improves the fit quality in the low $Q^2$ region (see the dashed curve at Fig. 1, left panel).
Indded, the maximal value of $\chi^2/$d.o.f. $\simeq 1.5$ at $Q^2_{min}=2$ GeV$^2$ is significantly lower then  $\chi^2/$d.o.f.
$\simeq 3$ of the DGLAP fit. However, it is important to stress that a truncation of the twist series (up to twist-6) is required
to have a good description of the data. Such truncation, however, may be motivated in QCD. Let
us recall that in BFKL, at the leading logarithmic approximation, only one reggeized gluon may couple to a fundamental
colour line. Since DGLAP and BFKL approximations have the same double logarithmic ($\ln x \ln Q^2$) limit, one concludes that also in DGLAP couplings of more than two gluons to a colour dipole is much weaker than in the eikonal picture. Thus one can
couple only 2 gluons to a colour dipole and up to four gluons to $q\bar{q}g$ component (two colour dipoles in the large $N_c$ limit)
without violating BFKL constraint. This means that one may expect a suppression beyond twist-8 if only the $q\bar{q}$ and $q\bar{q}g$
components are included in the calculations. This qualitative argument requires, obviously, further detailed studies.

\section{Acknowledgements}
The work is supported by the Polish National Science Centre grant no. DEC-2011/01/B/ST2/03643.


\begin{thebibliography}{99}
\bibitem{collins} J. Collins, Phys. Rev. \textbf{D57} (1998) 3051.
\bibitem{bartels} J. Bartels, K. Golec-Biernat and L. Motyka, Phys. Rev. \textbf{D81} (2010) 054017.
\bibitem{ZEUS} S. Chekanov et al. [ZEUS Collaboration], Nucl. Phys. \textbf{B831} (2010) 1; Nucl. Phys. \textbf{B816} (2009) 1.
\bibitem{MSS} L. Motyka, M. Sadzikowski and W. Slominski,  arXiv:hep-ph/1203.5461 (2012).
\bibitem{NZ} N. N. Nikolaev and B. G. Zakharov, Z. Phys. \textbf{C49} (1991) 607;
\bibitem{GBW} K. J. Golec-Biernat and M. W\"usthoff, Phys. Rev. \textbf{D59} (1998) 014017, Phys. Rev. \textbf{D60} (1999) 114023.
\bibitem{MKW} H.~Kowalski, L.~Motyka and G.~Watt, Phys. Rev.  \textbf{D74} (2006) 074016.
\bibitem{marquet} C. Marquet, Phys. Rev. \textbf{D76} (2007) 094017.
\bibitem{wusthoff} M. W\"usthoff, Phys. Rev. \textbf{D56} (1997) 4311.
\bibitem{munier_shoshi} S. Munier and A. Shoshi, Phys. Rev. \textbf{D69} (2004) 074022.
\end{thebibliography}
\end{document}